\documentclass[prd, aps, superscriptaddress, preprintnumbers, twocolumn, floatfix, nofootinbib]{revtex4}
\pdfoutput=1

\usepackage{amsfonts}
\usepackage{amsmath}
\usepackage{amssymb}
\usepackage{bm}
\usepackage{dcolumn}
\usepackage{graphicx}
\usepackage[latin1]{inputenc}
\usepackage{latexsym}
\usepackage{rotating}
\usepackage{hyperref}
\usepackage{graphicx}
\usepackage{color}

\newcommand\be{\begin{equation}}
\newcommand\ba{\begin{eqnarray}}
\newcommand\ee{\end{equation}}
\newcommand\ea{\end{eqnarray}}

\begin{document}

\title{Reheating after S-Brane Ekpyrosis}

\author{Robert Brandenberger}
\email{rhb@physics.mcgill.ca}
\affiliation{Department of Physics, McGill University, Montr\'{e}al, QC, H3A 2T8, Canada}

\author{Keshav Dasgupta}
\email{keshav@physics.mcgill.ca}
\affiliation{Department of Physics, McGill University, Montr\'{e}al, QC, H3A 2T8, Canada}

\author{Ziwei Wang}
\email{ziwei.wang@mail.mcgill.ca}
\affiliation{Department of Physics, McGill University, Montr\'{e}al, QC, H3A 2T8, Canada}

\date{\today}

\begin{abstract}
 
 In recent work, two of us proposed a nonsingular Ekpyrotic cosmology making use of an S-brane which forms at the end of the phase of Ekpyrotic contraction. This S-Brane mediates a transition between contraction and expansion. Graviitational waves passing through the S-Brane acquire a roughly scale-invariant spectrum, and if the S-Brane has zero shear, then a roughly scale-invariant spectrum of cosmological perturbatiions results. Here, we study the production of gauge field fluctuations driven by the decay of the S-Brane, and we show that
 the reheating process via gauge field production will be efficient, leading to a radiation-dominated expanding phase.
 
\end{abstract}

\pacs{98.80.Cq}
\maketitle

\section{Introduction}
\label{sec:intro}

Although the inflationary scenario \cite{Guth} has become the standard paradigm of early universe cosmology, it has recently been challenged from considerations based on fundamental physics. On one hand, inflation (at least canonical single field slow-roll inflation) seems hard to realize in string theory based on the {\it swampland constraints} \cite{swamp} (see e.g. \cite{swamprev} for a review). On the other hand, the effective field theory description of inflation is subject to the {\it Trans-Planckian Censorship} constraint (TCC) \cite{TCC} which forces the energy scale of inflation to be many orders of magnitude smaller than the scale of particle physics Grand Unification (the scale used in the canonical single field slow-roll models of inflation), leading to a negligible amplitude of primorial gravitational waves \cite{TCC2} (see, however, \cite{Dvali} for an opposite point of view on this issue).

Inflation is not the only early universe scenario which can explain the current observational data (see e.g. \cite{RHBalt} for a review of some alternative scenarios). Bouncing cosmologies (see e.g. \cite{Peter} for a recent review) and emergent scenarios such as {\it String Gas Cosmology} \cite{BV} are promising alternatives. However, at the moment none of the alternatives have been developed to the level of self-consistency that the inflationary scenario has. However, in light of the conceptual challenges to inflation, there is a great need to work on improvements of alternative scenarios.

Among bouncing cosmologies, the Ekpyrotic scenario \cite{Ekp} has a preferred status. The Ekpyrotic scenario is based on the hypothesis that the contracting phase was one of very slow contraction. Such a phase can be realized in the context of Einstein gravity by assuming that matter is dominated by a scalar field $\varphi$ with a negative exponential potential (but positive total energy density) such that the equation of state is $w \gg 1$, where $w = p / \rho$ is the ratio betwen pressure and energy density. As a consequence of this equation of state, all initial cold matter, radiation, spatial curvature and anisotropy is suppressed relative to the energy density in $\varphi$. Thus, the Ekpyrotic scenario solves the flatness problem of Standard Big Bang cosmology, and the homogeneous and isotropic contracting trajectory is an attractor in initial condition space, as first discussed in \cite{Erickson}. In the words of the recent paper \cite{Ijjas}, the Ekpyrotic scenario is a {\it super-smoother} \footnote{Note that the homogeneous and isotropic expanding phase in large field inflation is also a local attractor in initial condition space \cite{Kung}, as recently reviewed in \cite{RHBICrev}. Large field inflation is, however, in tension with the swampland criteria and the TCC \cite{TCC2}.}.

In order to yield a successful early universe cosmology, there must be a well-controlled transition from the contracting Ekpyrotic phase to the radiation phase of Standard Big Bang Cosmology. In the past, the transition was either assumed to be singular (breakdown of the effective field theory description \cite{Ekp}) or else obtained by invoking matter fields which violate the Null Energy Condition \cite{NewEkp} (see e.g \cite{Lehners} for a review). In a recent work, two of us have suggested \cite{Ziwei1} that the transition arises as a consequence of an S-brane which appears once the background energy density reaches the string scale \footnote{See \cite{Kounnas} for previous work on obtaining a non-singular bounce making use of an S-brane in a string theoretical setting.}. From the point of view of string theory, the S-brane represents string degrees of freedom which decouple at low energy densities but become comparable in energy to the energy scale of the background and hence have to be included in the low energy effective field theory as a space-filling object which is localized at a fixed time. Such a brane has vanishing energy energy density, and negative pressure (positive tension). As such, it can mediate the transition between the contracting phase and an expanding universe. In \cite{Ziwei1} it was shown that gravitational waves passing through the S-brane acquire a scale-invariant spectrum (with a slight blue tilt) if they begin as vacuum fluctuations early in the contracting phase. In \cite{Ziwei2} it was then shown that, assuming that the S-brane has zero shear, the curvature fluctuations similarly obtain a scale-invariant spectrum (this time with a slight red tilt), and two consistency relations which related the amplitudes and tilts of the scalar and tensor spectra were derived.

In the works \cite{Ziwei1, Ziwei2} it was assumed that the universe is radiation dominated after the transition. Here we study the production of radiation during the decay of the S-brane and show that, indeed, the post S-brane state is dominated by radiation. In analogy to the reheating which takes place at the end of inflation, where particle production is driven by the dynamics of the inflaton field \cite{DK, TB} (see \cite{RHrevs} for reviews), we can speak of the process of {\it reheating after Ekpyrosis} which we analyze here.

In the following section, we review the S-brane bounce mechanism and discuss the string theoretic setting of the S-brane. In Section 3 we then discuss the coupling of the S-brane to the Standard Model radiation field and then (in Section 4) estimate the efficiency of radiation production during the decay of the S-brane. We conclude with a summary and discussion of our results.

We assume a spatially flat homogeneous and isotropic background space-time with scale factor $a(t)$, where $t$ is physical time. It is often convenient to use conformal time $\tau$ determined by $dt = a(t) d\tau$. The Hubble expansion rate is $H \equiv {\dot{a}}/a$, and its inverse is the Hubble radius, the length scale which plays a key role in the evolution of cosmological fluctuations. We use units in which the speed of light and Planck's constant are set to $1$. Comoving spatial coordinates are denoted by ${\bf{x}}$, and the corresponding comoving momentum vector is ${\bf{k}}$ (its magnitude is written as $k$). The reduced Planck mass is denoted by $m_{pl}$ and it is related to Newton's gravitational constant $G$ via $m_{pl}^{-2} = 8 \pi G$.

\section{S-Brane Bounce}

The Ekpyrotic scenario is based on an effective field theory involving a canonically normalized scalar field $\varphi$ with negative exponential potential
\be \label{pot}
V(\varphi) \, = \, - V_0 e^{- \sqrt{2/p} \varphi / m_{pl}} \,
\ee
with $V_0 > 0$ and $0 < p \ll 1$ coupled to Einstein gravity. The universe is assumed to begin in a contracting phase at large values of $\varphi$ and positive total energy. The equation of state of the scalar field matter is
\be
w \, \equiv \, \frac{p}{\rho} \, = \, \frac{4}{3p} \, \gg \, 1 \, .
\ee
Hence, in the contracting phase, $\varphi$ comes to dominate over all other forms of energy which might have been present at the initial time.

Negative exponential potentials are ubiquitous in string theory. In fact, the first realization of the Ekpyrotic scenario \cite{Ekp} was based on the potential of a bulk brane moving in a Horava-Witten background \cite{HW}. In the context of string theory, we know that as the energy density of the background approaches the string scale, new stringy degrees of freedom become low mass and have to be included in the low energy effective action. In \cite{Ziwei1}, we proposed to do this by adding an S-brane \cite{Sbrane} to the action to yield
\ba \label{action}
S \, &=& \, \int d^4x \sqrt{-g} \left[ R + \frac{1}{2} \partial_{\mu} \varphi \partial^{\mu} \varphi
- V(\varphi) \right] \nonumber \\
& & - \int d^4x \kappa \delta(t - t_B) \sqrt{\gamma} \, .
\ea
The second term is the S-brane. It is localized at the time $t_B$ when the background density reaches the string scale. In the above, $R$ is the Ricci scalar and $g$ is the determinant of the four-dimensional space-time metric $g_{\mu \nu}$, $\gamma$ is the determinant of the induced metric on the S-brane world volume, and $\kappa$ is the tension of the S-brane, given by the string scale $\eta_s$.

As shown in \cite{Ziwei1}, the S-brane mediates a non-singular transition between Ekpyrotic contraction and expansion. Since the homogeneous and isotropic contracting solution is an attractor in initial condition space and any pre-existing classical fluctuations at the beginning of the phase of Ekpyrotic contraction get diluted relative to the contribution of $\varphi$, it is reasonable to assume that both scalar and tensor fluctuations are in their vacuum state on sub-Hubble scales. It was shown \cite{Ziwei1} the gravitational waves passing through the S-brane acquire a scale-invariant spectrum, and in \cite{Ziwei2} it was shown that, provided that the S-brane has zero shear,  a scale-invariant spectrum of curvature fluctuations is generated. The scenario leads in fact to two consistency relations between the four basic cosmological observables \cite{Ziwei2}, namely the amplitudes and tilts of the scalar and tensor spectra.

In \cite{Ziwei1, Ziwei2} it was simply assumed that the state after the cosmological bounce would be the radiation phase of Standard Big Bang cosmology, analogously to how the Standard Model radiation phase follows after inflation. In order for this to be the case, there has to be efficient energy transfer from the S-brane to radiation. In this paper, we show that this reheating process exists and is indeed very efficient.

In order to be able to study reheating after the S-brane bounce, we need to better understand the S-brane microphysics. Let us consider, to be specific, Type IIB superstring theory. In this theory, the gauge fields of the Standard Model of particle physics, including the photon field, live on the world volume of D-branes. In the context of Type IIB string theory, our four-dimensional space-time can be considered to be the world volume of a D3 brane (three spatial and one time dimension) which is located at a particular point along the compactified spatial dimensions.

Similar to what was done in the original Ekpyrotic proposal \cite{Ekp}, we assume that $\varphi$ is a K\"ahler modulus related to the size of one of the compact dimensions of space. Negative exponential potentials for such moduli fields are ubiquitous in string theory (see e.g. \cite{Baumann} for a review). In line with the setup of the Ekpyrotic scenario, the evolution begins in a phase of contraction with positive total energy density, the kinetic energy density being slightly larger than the potential energy density. As the universe contracts, the energy density increases until it reaches the string scale, which happens at a value of $\varphi$ which we denote by $\varphi_s$. At that point, the kinetic energy density of $\varphi$ is large enough to excite an S-brane.  Similar to what was done in \cite{Sbrane}, we will model the S-brane by a tachyon condensate. The tachyon configuration $T(x, t)$ can be viewed as an unstable four brane, the extra spatial dimension being, for example, the dimension corresponding to $\varphi$.  Since anisotropies and matter inhomogeneities are smoothed out in the contracting phase, the tachyon configuration will be homogeneous $T(x, t) = T(t)$ on the three brane of our space-time volume.

Note that the potential of $\varphi$ at $\varphi_s$ is negative. Its value can be viewed as the magnitude of the (negative) cosmological constant in the anti-de-Sitter ground state of the string theory. The tachyon configuration can be described by a positive potential energy density $V(T)$ on top of the negative value of the background. We take the potential to have the form
\begin{equation}\label{potel}
 V(T) =
 \begin{cases}
  -\frac{1}{2}\lambda \eta^2 T^2 + \frac{1}{2}\lambda \eta^4, ~~~-\eta < T < \eta \\
  0, ~~~|T|>\eta     \, ,                                                         
 \end{cases}
\end{equation}
where we expect $\eta$ to be given by the string scale. A sketch of this potential is given in Fig. 1.
In order to represent an S-brane, the total energy density $V_{\rm{grav}}$ (tachyon energy plus contribution from the cosmological constant) must vanish at $T = 0$. This forces $\eta$ to be of the order of the string energy scale. It is the quantity $V_{\rm{grav}}$ which determines the evolution of the gravitational background.

The action of a Standard Model gauge field (e.g. the photon field) on the D3 brane has the Born-Infeld form (see e.g. \cite{Minahan})
\be
S_{BI} \, = \, \int dt d^3x V_{\rm{total}}(T) \sqrt{{\rm{det}}(\eta_{\mu \nu} + F_{\mu \nu})} \, ,
\ee
where $\tau$ is the tension of the D-brane, $V_{\rm{total}}(T)$ is the total potential energy of the tachyon, and $F_{\mu \nu}$ is the field strength of the gauge field $A_{\mu}$ (in string units). The total potential energy gets a constant contribution from the tension $\tau$ of the D3-brane plus the contribution from the tachyon field:
\be
V_{\rm{total}}(T) \, = \, \tau + V(T) + f(\partial_0 T, \partial_0^2 T, ...)\, .
\ee
where $f(\partial_0 T, \partial_0^2 T, ...)$ can be ignored for our case \cite{senbps}, and
the constant term coming from the brane tension does not have any effect on the particle production calculation which we discuss below.

We are interested in the growth of fluctuations of $A_{\mu}$ which start out in their vacuum state. Hence, we can expand the square root and keep only the leading term in $F_{\mu \nu}F^{\mu \nu}$
\be
\sqrt{{\rm{det}}(\eta_{\mu \nu} + F_{\mu \nu})} \,
\simeq \, 1 + \frac{1}{4} F_{\mu \nu}F^{\mu \nu} \, .
\ee
In particular, we then recover the usual gauge theory action for $|T| > \eta$ where $V_{\rm{total}}(T)$ is a non-vanishing constant.

If we choose the gauge $A_0 = 0$, then for a homogeneous tachyon kink $K(t)$ the action for the gauge fields becomes \cite{Minahan}
\be
S \, = -\, \int dt d^3x {\dot{K}}^2(t)
\left(\frac{1}{4} F_{ab}F^{ab} - \frac{1}{2} {\dot{A_a}}{\dot{A^a}} \right) \, ,
\ee
where the indices $a, b$ are spatial ones, and we have extracted the terms with time derivatives. We have also used the on-shell condition in (\ref{potel}) to replace $V_{\rm total}$ by $\dot{K}(t)$, ignoring constant factors. Note that,
during passage through the S-brane, the coefficient ${\dot{K}}$ depends on time, and hence gauge field particle production is possible. This is what we study in the following two sections.

\section{Coupling of Radiation to the S-Brane}

We have seen that the coupling of the S-brane to the gauge field $A_{\mu}$ is given by \cite{Minahan}
\be \label{Aaction}
S \, = \, \int dt d^3x {\dot{K}}^2(t) \left( \frac{1}{4} F_{ab} F^{ab} - \frac{1}{2} {\dot{A}}_a {\dot{A}}^a  \right) \, ,
\ee
where the indices $a, b$ run only over space, and we have used temporal gauge. This action can be put into canonical form by defining a rescaled gauge field
\be
B_a \, \equiv \, {\dot{K}} A_a
\ee
with associated field strength tensor ${\tilde{F}}_{ab}$. In terms of this rescaled gauge field, the action is given by \cite{Minahan}
\be \label{Baction}
S \, = \, \int dt d^3x \left[ \frac{1}{4} {\tilde{F}}_{ab} {\tilde{F}}^{ab}
+ \frac{1}{2} B_{a} \left( - \frac{\partial^2}{\partial^2t} + \frac{K_{,ttt}}{K_{,t}}\right) B^{a} \right] \, ,
\ee
where  $K_{,ttt}$ stands for the third time derivative of $K$.

The equation of motion for $T$ in the range $-\eta < T < \eta$ is
\be
{\ddot{T}} - \lambda \eta^2 T \, = \, 0 \, ,
\ee
and its solution is
\begin{equation}
 T(t) =
 \begin{cases}
  v \sqrt{\frac{\eta ^4 \lambda }{v^2}+1}(t+P)-\eta, ~~~ t<-P          \\
  \frac{v\sinh (\eta\sqrt{\lambda}t)}{\eta\sqrt{\lambda}}, ~~~ -P<t<P  \\
  v \sqrt{\frac{\eta ^4 \lambda }{v^2}+1}(t-P)+\eta, ~~~ t>P      \, , 
 \end{cases}
 \label{eq: kink solt}
\end{equation}
where we have adjusted the time axis such that $T$ is at the top of the potential at time $t = 0$, and $v$ is the velocity of the field at the top. The time interval $P$ is the duration of the bounce, and is determined by when $T(P) = \eta$.

The scenario we have in mind is now the following. During the phase of Ekpyrotic contraction, $\varphi$ is decreasing from a very large initial positive value. Once the energy density of $\varphi$ approaches the string density the S-brane forms. The formation of the S-brane is the process where $T$ increases from $T = - \eta$ to $T = 0$. The decay of the S-brane corresponds to the motion of $T$ from $T = 0$ to $T = \eta$. During the phase of Ekpyrotic contraction, the total energy density in the $\varphi$ field is slightly positive. This implies that the magnitude of the kinetic energy is close to (but larger) than the absolute value of the total potential energy $V_{\rm{total}}$. Hence we expect
\be
v \, \simeq \, \lambda^{1/2} \eta^2 \, ,
\ee
where $v$ is defined in (\ref{eq: kink solt}). This implies that
\be
P \, \simeq \, \lambda^{-1/2} \eta^{-1} \, .
\ee
To make contact with the discussion in \cite{Minahan}), we note that in our case the profile function $K(t)$ is given by the soliton solution of the equation of motion for $T(t)$:
\be \label{soliton}
K(t) \, = \, T(t) \, ,
\ee
which implies that
\be
\frac{K_{,ttt}}{K_{,t}} \, = \, \lambda \eta^2 \, .
\ee

\begin{figure}
 \includegraphics[width = 0.45\textwidth]{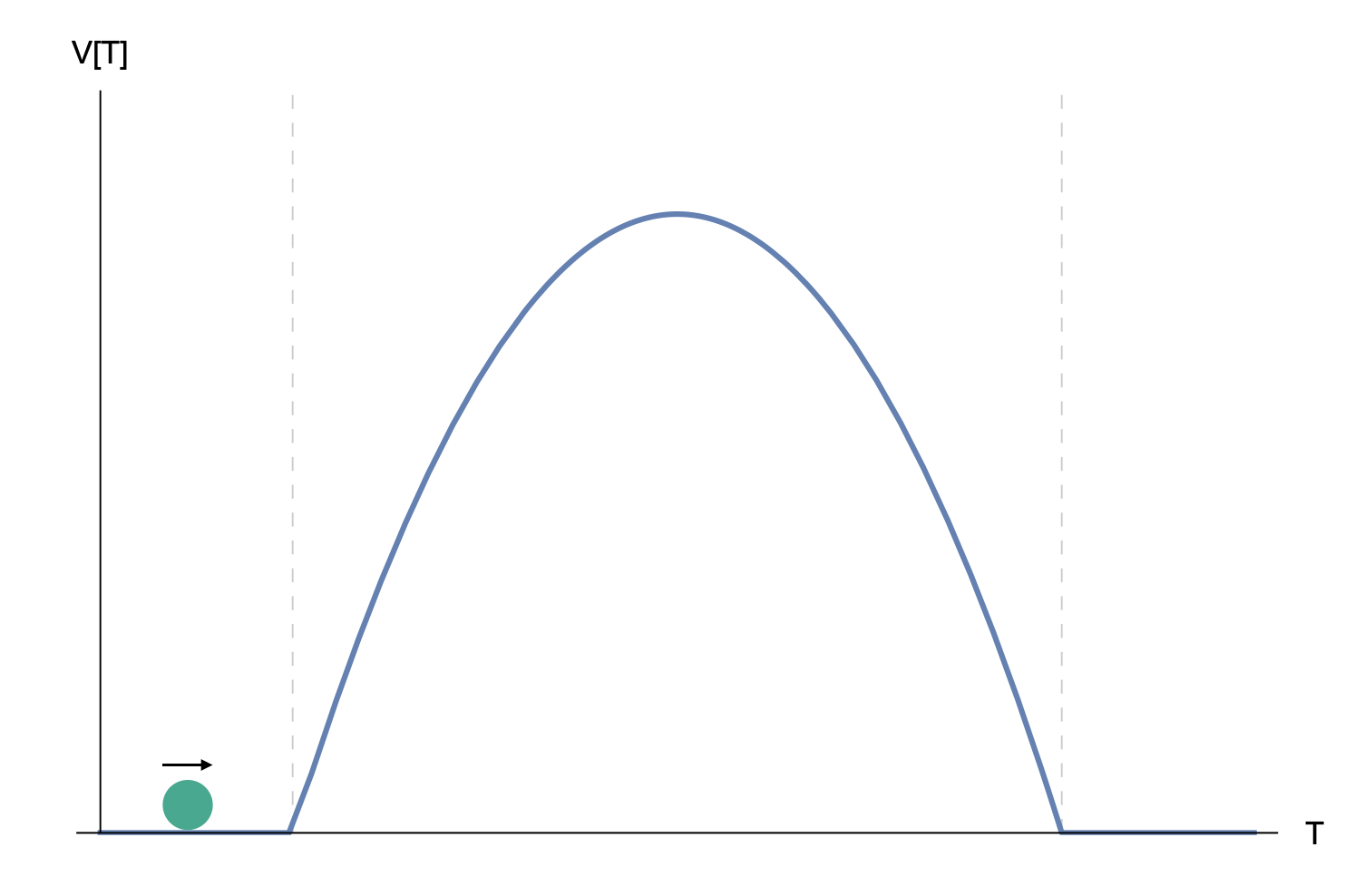}
 \caption{The sketch of tachyon potential $V(T)$ during the bounce phase.}
 \label{fig:potential}
\end{figure}

\section{Reheating from S-Brane Decay}

As discussed in the previous section, the coupling of the rescaled gauge field $B_{\mu}$ to the S-brane profile function $K(t)$ is given by (\ref{Baction}). The resulting equation of motion for the gauge field $B_i$ is
\be
{\ddot{B_i}} - \partial_a\partial^a B_i - \lambda \eta^2 B_i \, = \, 0 \, ,
\ee
which is a wave equation with tachyonic mass term. The solution for the Fourier modes is given by
\ba
B_i(k,t) \, &=& \, c_1 e^{t \sqrt{\eta ^2 \lambda -k^2}}+c_2 e^{-t \sqrt{\eta ^2 \lambda -k^2}}, ~~~~ k < k_c  \\
B_i(k,t) \, &=& \, c_1 e^{ i t \sqrt{\eta ^2 \lambda -k^2}}+c_2 e^{- i t \sqrt{\eta ^2 \lambda -k^2}}, ~~~~ k > k_c   \nonumber  \, ,
\label{eq: sol B}
\ea
with $k_c \equiv \lambda^{1/2} \eta$. From (\ref{eq: sol B}) it is not hard to see that the growing mode $k\rightarrow 0$ for $t>0$ asymptotically scales as $e^{\eta\sqrt{\lambda}t}$. The corresponding modes of $A_i$ field, however, is a constant mode. Therefore there is no growing mode for $A_i$ after $t = 0$.

In terms of the canonical gauge field, the equation of motion becomes (inserting the expression (\ref{soliton}) for $K(t)$)
\begin{equation}
 \ddot A_i + 2\eta  \sqrt{\lambda } \tanh \left(\eta  \sqrt{\lambda } t\right) \dot A_i + k^2 A_i \, = \,  0 \, .
 \label{eq: EoM of A}
\end{equation}
The solution of this equation can be written as follows
\begin{widetext}
 \begin{equation}
  A_i(t,k) = \text{sech}\left(\eta  \sqrt{\lambda } t\right) \times
  \begin{cases}
   \tilde c_1 e^{-t \sqrt{\eta ^2 \lambda                                                    
   -k^2}}+\tilde c_2 e^{t \sqrt{\eta ^2 \lambda -k^2}} , ~~~~ k < k_c = \eta\sqrt{\lambda}   \\
   \tilde c_1 e^{-it \sqrt{-\eta ^2 \lambda  +                                               
   k^2}}+ \tilde c_2 e^{it \sqrt{-\eta ^2 \lambda +k^2}} , ~~~~ k > k_c = \eta\sqrt{\lambda} 
  \end{cases}
  \label{eq: gfmodes cros}
 \end{equation}
 In the asymptotic regions $t\rightarrow \pm\infty $, the equation \eqref{eq: EoM of A} has solutions
 \begin{equation}
  A_i(k,t\rightarrow \infty) \sim e^{-\eta\sqrt{\lambda}t} \times
  \begin{cases}
   e^{\pm\sqrt{\eta^2\lambda - k^2}t}, ~~~~ k < k_c = \eta\sqrt{\lambda}       \\
   e^{\pm i (-\sqrt{\eta^2\lambda + k^2})t}, ~~~~ k > k_c = \eta\sqrt{\lambda} 
  \end{cases}
 \end{equation}
 and
 \begin{equation}
  A_i(k,t\rightarrow -\infty) \sim e^{+\eta\sqrt{\lambda}t} \times
  \begin{cases}
   e^{\pm\sqrt{\eta^2\lambda - k^2}t}, ~~~~ k < k_c = \eta\sqrt{\lambda}            \\
   e^{\pm i (-\sqrt{\eta^2\lambda + k^2})t}, ~~~~ k > k_c = \eta\sqrt{\lambda} \, . 
  \end{cases}
 \end{equation}
\end{widetext}

During the phase of Ekpyrotic contraction, any pre-existing gauge field fluctuations are diluted compared to the energy density in $\varphi$. Hence, the gauge field fluctuations will be in their ground state when $\varphi$ hits the S-brane. Hence, focusing on modes with $k < k_c$, we can write the solution for $A_i$ in the form
\begin{widetext}
 \begin{equation}
  A_i =
  \begin{cases}
   A_i^{(1)} = \frac{e^{-i k t}}{\sqrt{2} \sqrt{k}} ~~~~ t < -P                                                       \\
   A_i^{(2)} = \text{sech}\left(\eta  \sqrt{\lambda } t\right)\left( c_1 e^{-t \sqrt{\eta ^2 \lambda                  
   -k^2}}+  c_2 e^{t \sqrt{\eta ^2 \lambda -k^2}} \right )  ~~~~ -P<t<P                                               \\
   A_i^{(3)} = \alpha\frac{e^{-i k t}}{\sqrt{2} \sqrt{k}} + \beta \frac{e^{+i k t}}{\sqrt{2} \sqrt{k}} ~~~~ t >P \, , 
  \end{cases}
  \label{eq: modes all}
 \end{equation}
\end{widetext}
where the constants $c_1$, $c_2$, $\alpha$ and $\beta$ (which are all functions of $k$) are determined by matching $A_i$ and ${\dot{A}}_i$ at $t = -P$ and $t = P$, i.e. using the junction conditions
\begin{equation}
 \begin{cases}
  A_i^{(1)}(-P) = A_i^{(2)}(-P), ~~~   \dot A_i^{(1)}(-P) = \dot A_i^{(2)}(-P)   \\
  A_i^{(2)}(P) = A_i^{(3)}(P), ~~~   \dot A_i^{(2)}(P) = \dot A_i^{(3)}(P)  \, . 
  \label{eq: junction}                                                           
 \end{cases}
\end{equation}
Using \eqref{eq: modes all} and \eqref{eq: junction} we can solve for the Bogoliubov mode mixing coefficients $\beta$. Notice that $|\beta|^2 - |\alpha|^2 = 1$ by the requirement of unitarity. The result is
\ba
|\beta|^2 &=& \frac{\mu^2\text{sech}^4(\mu  P)}{16k^2(1 - \frac{k^2}{\mu^2})} \\ && \Bigg[
\left(\sqrt{1-\frac{k^2}{\mu^2}}+ 1\right) \sinh \left(2\mu P\left(1 -\sqrt{1-\frac{k^2}{\mu^2}}\right)\right)\nonumber \\
&+& \left(\sqrt{1-\frac{k^2}{\mu^2}}-1\right) \sinh \left(2\mu P
\left(\sqrt{1-\frac{k^2}{\mu^2}}+1 \right)\right)\Bigg]^2 \nonumber
\label{eq: pn}
\ea
where we introduced the constant $\mu \equiv \sqrt{\lambda}\eta$ to simplify the expression. In the infrared limit $k \ll k_c$ we obtain
\be
|\beta|^2 \, \approx \, \frac{k^2\text{sech}^4(\mu P)\Big(4P\mu +\text{sinh}(4P\mu)\Big)^2}{64\mu^2\left(1 - \frac{k^2}{\mu^2}\right)} \, \sim \, k^2 \, ,
\ee
while for $k \sim k_c$ we get
\be \label{upper}
|\beta|^2 \, \approx  \, -\frac{\mu ^2 \text{sech}^4(\mu  P) (\sinh (2 \mu  P)-2
 \mu  P \cosh (2 \mu  P))^2}{4 k^2}
\ee
Note that for $k > k_c$ the Bogoliubov coefficients are suppressed.

Since the particle number density is given by the Bogoliubov mode mixing coefficient, i.e. $n(k) \sim |\beta|^2$, we can use the above results to give an order of magnitude estimate of the energy density in gauge fields produced during S-brane decay. The energy density in the gauge fields can be obtained by integrating over the contribution of all modes with $k < k_c$, making use of the initial vacuum value
\be
A_k(0)  \, = \, \frac{1}{\sqrt{2k}} \, .
\ee
Making use of the solution \eqref{eq: modes all} we obtain
\be
\rho_A(P) \, \sim \, \int_0^{k_c} dk \frac{k^2}{2\pi^2} \frac{1}{2k} |\beta_k|^2 k^2 \, \sim \, \frac{1}{32\pi} \lambda^2 \eta^4 \, .
\ee
Here, the first factor of $k^2$ comes from the phase space measure, and the second factor of $k^2$ comes from the gradients (recall that the energy is gradient energy). The integral is dominated by the upper limit, and, inserting the value for $k_c$ and the result (\ref{upper}) for the Bogoliubov coefficients, we obtain
\be \label{mainres}
\frac{\rho_A(\tau)}{V_0} \, \sim \, \frac{3}{16\pi} \lambda \, .
\ee
This is the main result of our analysis. We expect $\lambda \sim 1$, and hence we see that the S-brane can very efficiently decay into radiation.

\section{Conclusions and Discussion} \label{conclusion}

We have studied the embedding of the S-brane bounce scenario proposed in \cite{Ziwei1} into string theory. Considering Type IIB superstring theory, our space-time can be viewed as a D3-brane on which the usual Standard Model fields (including the gauge field of electromagnetism) live. At the end point of a phase of Ekpyrotic contraction, an unstable D4-brane is excited and acts gravitationally as an S-brane. This tachyonic brane fills our three spatial dimensions and is extended in one further direction (e.g. the direction which corresponds to the Ekpyrotic scalar field $\varphi$). The tachyon configuration couples to the gauge fields via a Born-Infeld action. Hence, during the formation and decay of the D4-brane, gauge field production can occur. We have shown that the process of gauge field production is very effective and can transfer a fraction of order one of the S-brane energy into radiation. Hence, there is no obstruction to efficient reheating after the S-brane bounce. The S-brane will automatically mediate not only the transition between an initial contracting phase and an expanding phase, but in fact to the expanding radiation phase of Standard Big Bang cosmology.

\section*{Acknowledgement}

\noindent The research at McGill is supported in part by funds from NSERC and from the Canada Research Chair program. RB is grateful for hospitality of the Institute for Theoretical Physics and the Institute for Particle Physics and Astrophysics of the ETH Zurich during the completion of this project. ZW acknowledges partial support from a McGill Space Institute Graduate Fellowship and from a Templeton Foundation Grant.


\begin{thebibliography}{99}
 
 \bibitem{Guth}
 A.~H.~Guth,
 ``The Inflationary Universe: A Possible Solution to the Horizon and Flatness Problems,''
 Phys.\ Rev.\ D {\bf 23}, 347 (1981)
 [Adv.\ Ser.\ Astrophys.\ Cosmol.\  {\bf 3}, 139 (1987)].
 doi:10.1103/PhysRevD.23.347;\\
 R.~Brout, F.~Englert and E.~Gunzig,
 ``The Creation Of The Universe As A Quantum Phenomenon,''
 Annals Phys.\  {\bf 115}, 78 (1978);\\
 A.~A.~Starobinsky,
 ``A New Type Of Isotropic Cosmological Models Without Singularity,''
 Phys.\ Lett.\ B {\bf 91}, 99 (1980);\\
 K.~Sato,
 ``First Order Phase Transition Of A Vacuum And Expansion Of The Universe,''
 Mon.\ Not.\ Roy.\ Astron.\ Soc.\  {\bf 195}, 467 (1981).
 
 \bibitem{swamp}
 H.~Ooguri and C.~Vafa,
 ``On the Geometry of the String Landscape and the Swampland,''
 Nucl.\ Phys.\ B {\bf 766}, 21 (2007)
 doi:10.1016/j.nuclphysb.2006.10.033
 [hep-th/0605264];\\
 G.~Obied, H.~Ooguri, L.~Spodyneiko and C.~Vafa,
 ``De Sitter Space and the Swampland,''
 arXiv:1806.08362 [hep-th];\\
 P.~Agrawal, G.~Obied, P.~J.~Steinhardt and C.~Vafa,
 ``On the Cosmological Implications of the String Swampland,''
 Phys.\ Lett.\ B {\bf 784}, 271 (2018)
 doi:10.1016/j.physletb.2018.07.040
 [arXiv:1806.09718 [hep-th]];\\
 S.~K.~Garg and C.~Krishnan,
  ``Bounds on Slow Roll and the de Sitter Swampland,''
  JHEP {\bf 1911}, 075 (2019)
  doi:10.1007/JHEP11(2019)075
  [arXiv:1807.05193 [hep-th]].
  
 \bibitem{swamprev}
 T.~D.~Brennan, F.~Carta and C.~Vafa,
 ``The String Landscape, the Swampland, and the Missing Corner,''
 PoS TASI {\bf 2017}, 015 (2017)
 doi:10.22323/1.305.0015
 [arXiv:1711.00864 [hep-th]];\\
 E.~Palti,
 ``The Swampland: Introduction and Review,''
 arXiv:1903.06239 [hep-th].
 
 \bibitem{TCC}
 A.~Bedroya and C.~Vafa,
 ``Trans-Planckian Censorship and the Swampland,''
 arXiv:1909.11063 [hep-th].
 
 \bibitem{TCC2}
 A.~Bedroya, R.~Brandenberger, M.~Loverde and C.~Vafa,
 ``Trans-Planckian Censorship and Inflationary Cosmology,''
 Phys.\ Rev.\ D {\bf 101}, no. 10, 103502 (2020)
 doi:10.1103/PhysRevD.101.103502
 [arXiv:1909.11106 [hep-th]].
 
 \bibitem{Dvali}
 G.~Dvali, A.~Kehagias and A.~Riotto,
 ``Inflation and Decoupling,''
 arXiv:2005.05146 [hep-th].
 
 \bibitem{RHBalt}
 R.~H.~Brandenberger,
 ``Cosmology of the Very Early Universe,''
 AIP Conf.\ Proc.\  {\bf 1268}, 3 (2010)
 doi:10.1063/1.3483879
 [arXiv:1003.1745 [hep-th]];\\
 R.~H.~Brandenberger,
 ``Alternatives to the inflationary paradigm of structure formation,''
 Int.\ J.\ Mod.\ Phys.\ Conf.\ Ser.\  {\bf 01}, 67 (2011)
 doi:10.1142/S2010194511000109
 [arXiv:0902.4731 [hep-th]].
 
 \bibitem{Peter}
 R.~Brandenberger and P.~Peter,
 ``Bouncing Cosmologies: Progress and Problems,''
 Found.\ Phys.\  {\bf 47}, no. 6, 797 (2017)
 doi:10.1007/s10701-016-0057-0
 [arXiv:1603.05834 [hep-th]].
 
 \bibitem{BV}
 R.~H.~Brandenberger and C.~Vafa,
 ``Superstrings In The Early Universe,''
 Nucl.\ Phys.\ B {\bf 316}, 391 (1989);\\
 A.~Nayeri, R.~H.~Brandenberger and C.~Vafa,
 ``Producing a scale-invariant spectrum of perturbations in a Hagedorn phase of string cosmology,''
 Phys.\ Rev.\ Lett.\  {\bf 97}, 021302 (2006)
 doi:10.1103/PhysRevLett.97.021302
 [hep-th/0511140].
 
 \bibitem{Ekp}
 J.~Khoury, B.~A.~Ovrut, P.~J.~Steinhardt and N.~Turok,
 ``The Ekpyrotic universe: Colliding branes and the origin of the hot big
 bang,''
 Phys.\ Rev.\ D {\bf 64}, 123522 (2001) [hep-th/0103239];\\
 J.~Khoury, B.~A.~Ovrut, N.~Seiberg, P.~J.~Steinhardt and N.~Turok,
 ``From big crunch to big bang,''
 Phys.\ Rev.\ D {\bf 65}, 086007 (2002)
 doi:10.1103/PhysRevD.65.086007
 [hep-th/0108187].
 
 \bibitem{Erickson}
 J.~K.~Erickson, D.~H.~Wesley, P.~J.~Steinhardt and N.~Turok,
 ``Kasner and mixmaster behavior in universes with equation of state w >= 1,''
 Phys.\ Rev.\ D {\bf 69}, 063514 (2004)
 doi:10.1103/PhysRevD.69.063514
 [hep-th/0312009].
 
 \bibitem{Ijjas}
 W.~G.~Cook, I.~A.~Glushchenko, A.~Ijjas, F.~Pretorius and P.~J.~Steinhardt,
 ``Supersmoothing through Slow Contraction,''
 arXiv:2006.01172 [gr-qc].
 
 \bibitem{Kung}
 R.~H.~Brandenberger and J.~H.~Kung,
 ``Chaotic Inflation as an Attractor in Initial Condition Space,''
 Phys.\ Rev.\ D {\bf 42}, 1008 (1990).
 doi:10.1103/PhysRevD.42.1008;\\
 H.~A.~Feldman and R.~H.~Brandenberger,
 ``Chaotic Inflation With Metric and Matter Perturbations,''
 Phys.\ Lett.\ B {\bf 227}, 359 (1989).
 doi:10.1016/0370-2693(89)90944-1;\\
 W.~E.~East, M.~Kleban, A.~Linde and L.~Senatore,
 ``Beginning inflation in an inhomogeneous universe,''
 JCAP {\bf 1609}, 010 (2016)
 doi:10.1088/1475-7516/2016/09/010
 [arXiv:1511.05143 [hep-th]];\\
 K.~Clough, E.~A.~Lim, B.~S.~DiNunno, W.~Fischler, R.~Flauger and S.~Paban,
 ``Robustness of Inflation to Inhomogeneous Initial Conditions,''
 JCAP {\bf 1709}, 025 (2017)
 doi:10.1088/1475-7516/2017/09/025
 [arXiv:1608.04408 [hep-th]].
 
 \bibitem{RHBICrev}
 R.~Brandenberger,
 ``Initial conditions for inflation -- A short review,''
 Int.\ J.\ Mod.\ Phys.\ D {\bf 26}, no. 01, 1740002 (2016)
 doi:10.1142/S0218271817400028
 [arXiv:1601.01918 [hep-th]].
 
 \bibitem{NewEkp}
 A.~Notari and A.~Riotto,
 ``Isocurvature perturbations in the ekpyrotic universe,''
 Nucl.\ Phys.\ B {\bf 644}, 371 (2002)
 doi:10.1016/S0550-3213(02)00765-4
 [hep-th/0205019];\\
 F.~Finelli,
 ``Assisted contraction,''
 Phys.\ Lett.\ B {\bf 545}, 1 (2002)
 doi:10.1016/S0370-2693(02)02554-6
 [hep-th/0206112];\\
 F.~Di Marco, F.~Finelli and R.~Brandenberger,
 ``Adiabatic and isocurvature perturbations for multifield generalized Einstein models,''
 Phys.\ Rev.\ D {\bf 67}, 063512 (2003)
 doi:10.1103/PhysRevD.67.063512
 [astro-ph/0211276];\\
 J.~L.~Lehners, P.~McFadden, N.~Turok and P.~J.~Steinhardt,
 ``Generating ekpyrotic curvature perturbations before the big bang,''
 Phys.\ Rev.\ D {\bf 76}, 103501 (2007)
 doi:10.1103/PhysRevD.76.103501
 [hep-th/0702153 [HEP-TH]];\\
 E.~I.~Buchbinder, J.~Khoury and B.~A.~Ovrut,
 ``New Ekpyrotic cosmology,'' Phys.\ Rev.\ D {\bf 76}, 123503 (2007)
 doi:10.1103/PhysRevD.76.123503 [hep-th/0702154];\\
 P.~Creminelli and L.~Senatore,
 ``A Smooth bouncing cosmology with scale invariant spectrum,''
 JCAP {\bf 0711}, 010 (2007)
 doi:10.1088/1475-7516/2007/11/010
 [hep-th/0702165].
 
 \bibitem{Lehners}
 J.~L.~Lehners,
 ``Ekpyrotic and Cyclic Cosmology,''
 Phys.\ Rept.\  {\bf 465}, 223 (2008)
 doi:10.1016/j.physrep.2008.06.001
 [arXiv:0806.1245 [astro-ph]].
 
 \bibitem{Ziwei1}
 R.~Brandenberger and Z.~Wang,
 ``Nonsingular Ekpyrotic Cosmology with a Nearly Scale-Invariant Spectrum of Cosmological Perturbations and Gravitational Waves,''
 Phys.\ Rev.\ D {\bf 101}, no. 6, 063522 (2020)
 doi:10.1103/PhysRevD.101.063522
 [arXiv:2001.00638 [hep-th]].
 
 \bibitem{Kounnas}
 R.~H.~Brandenberger, C.~Kounnas, H.~Partouche, S.~P.~Patil and N.~Toumbas,
 ``Cosmological Perturbations Across an S-brane,''
 JCAP {\bf 1403}, 015 (2014)
 doi:10.1088/1475-7516/2014/03/015
 [arXiv:1312.2524 [hep-th]];\\
 C.~Kounnas, H.~Partouche and N.~Toumbas,
 ``S-brane to thermal non-singular string cosmology,''
 Class.\ Quant.\ Grav.\  {\bf 29}, 095014 (2012)
 doi:10.1088/0264-9381/29/9/095014
 [arXiv:1111.5816 [hep-th]];\\
 C.~Kounnas, H.~Partouche and N.~Toumbas,
 ``Thermal duality and non-singular cosmology in d-dimensional superstrings,''
 Nucl.\ Phys.\ B {\bf 855}, 280 (2012)
 doi:10.1016/j.nuclphysb.2011.10.010
 [arXiv:1106.0946 [hep-th]].
 
 \bibitem{Ziwei2}
 R.~Brandenberger and Z.~Wang,
 ``Ekpyrotic Cosmology with a Zero-Shear S-Brane,''
 arXiv:2004.06437 [hep-th].
 
 \bibitem{DK}
 A.~D.~Dolgov and D.~P.~Kirilova,
 ``On Particle Creation By A Time Dependent Scalar Field,''
 Sov.\ J.\ Nucl.\ Phys.\  {\bf 51}, 172 (1990)
 [Yad.\ Fiz.\  {\bf 51}, 273 (1990)].
 
 \bibitem{TB}
 J.~H.~Traschen and R.~H.~Brandenberger,
 ``Particle Production During Out-of-equilibrium Phase Transitions,''
 Phys.\ Rev.\ D {\bf 42}, 2491 (1990).
 doi:10.1103/PhysRevD.42.2491
 
 \bibitem{RHrevs}
 R.~Allahverdi, R.~Brandenberger, F.~Y.~Cyr-Racine and A.~Mazumdar,
 ``Reheating in Inflationary Cosmology: Theory and Applications,''
 Ann.\ Rev.\ Nucl.\ Part.\ Sci.\  {\bf 60}, 27 (2010)
 doi:10.1146/annurev.nucl.012809.104511
 [arXiv:1001.2600 [hep-th]];\\
 M.~A.~Amin, M.~P.~Hertzberg, D.~I.~Kaiser and J.~Karouby,
 ``Nonperturbative Dynamics Of Reheating After Inflation: A Review,''
 Int.\ J.\ Mod.\ Phys.\ D {\bf 24}, 1530003 (2014)
 doi:10.1142/S0218271815300037
 [arXiv:1410.3808 [hep-ph]].
 
 \bibitem{HW}
 P.~Horava and E.~Witten,
 ``Eleven-dimensional supergravity on a manifold with boundary,''
 Nucl.\ Phys.\ B {\bf 475}, 94 (1996)
 doi:10.1016/0550-3213(96)00308-2
 [hep-th/9603142];\\
 P.~Horava and E.~Witten,
 ``Heterotic and type I string dynamics from eleven-dimensions,''
 Nucl.\ Phys.\ B {\bf 460}, 506 (1996)
 doi:10.1016/0550-3213(95)00621-4
 [hep-th/9510209].
 
 \bibitem{Sbrane}
 M.~Gutperle and A.~Strominger,
 ``Space - like branes,''
 JHEP {\bf 0204}, 018 (2002)
 doi:10.1088/1126-6708/2002/04/018
 [hep-th/0202210].
 
 \bibitem{Baumann}
 D.~Baumann and L.~McAllister,
 ``Inflation and String Theory,''
 doi:10.1017/CBO9781316105733
 arXiv:1404.2601 [hep-th].
 
 \bibitem{Minahan}
 J.~A.~Minahan and B.~Zwiebach,
 ``Gauge fields and fermions in tachyon effective field theories,''
 JHEP {\bf 0102}, 034 (2001)
 doi:10.1088/1126-6708/2001/02/034
 [hep-th/0011226].
 
 \bibitem{senbps}
 A.~Sen,
 ``Supersymmetric world volume action for nonBPS D-branes,''
 JHEP \textbf{10}, 008 (1999)
 doi:10.1088/1126-6708/1999/10/008
 [arXiv:hep-th/9909062 [hep-th]];\\
 A. Sen,
 ``Non-BPS D-branes in string theory,''
 Class. Quant. Grav. \textbf{17}, 1251-1256 (2000)
 doi:10.1088/0264-9381/17/5/334
 
 
\end{thebibliography}
\end{document}